\documentclass[pra,superscriptaddress,onecolumn, 11pt]{revtex4-2}
\usepackage{hyperref}
\usepackage{graphicx}
\usepackage{amsmath}
\usepackage{amssymb}
\usepackage{amsfonts}
\usepackage{latexsym,epsfig,bbm}

\usepackage{color}
\definecolor{blue}{rgb}{0,0.2,1}

\definecolor{red}{rgb}{0.9,0,0}

\newcommand{\bra}[1]{\langle{#1}|}
\newcommand{\ket}[1]{|{#1}\rangle}

\newcommand{\figref}[1]{Fig.~\ref{#1}}

\newcommand{\past}[1]{\overleftarrow{#1}}
\newcommand{\fut}[1]{\overrightarrow{#1}}
\newcommand{\pastfut}[1]{\overleftrightarrow{#1}}

\begin{document}

 \title{Quantum-inspired identification of complex cellular automata}

\author{Matthew Ho}
\email{matthew.ho.291@gmail.com}
\affiliation{Institute of High Performance Computing (IHPC), Agency for Science, Technology and Research (A*STAR), 1 Fusionopolis Way, \#16-16 Connexis, Singapore 138632, Republic of Singapore}
\affiliation{Nanyang Quantum Hub, School of Physical and Mathematical Sciences, Nanyang Technological University, Singapore 637371, Singapore}
\affiliation{Complexity Institute, Nanyang Technological University, Singapore 637335, Singapore}

\author{Andri Pradana}
\email{andr0071@e.ntu.edu.sg}
\affiliation{Nanyang Quantum Hub, School of Physical and Mathematical Sciences, Nanyang Technological University, Singapore 637371, Singapore}

\author{\mbox{Thomas J.~Elliott}}
\email{physics@tjelliott.net}
\affiliation{Department of Physics \& Astronomy, University of Manchester, Manchester M13 9PL, United Kingdom}
\affiliation{Department of Mathematics, University of Manchester, Manchester M13 9PL, United Kingdom}
\affiliation{Department of Mathematics, Imperial College London, London SW7 2AZ, United Kingdom}
\affiliation{Complexity Institute, Nanyang Technological University, Singapore 637335, Singapore}
\affiliation{Nanyang Quantum Hub, School of Physical and Mathematical Sciences, Nanyang Technological University, Singapore 637371, Singapore}

\author{Lock Yue Chew}
\email{lockyue@ntu.edu.sg}
\affiliation{Nanyang Quantum Hub, School of Physical and Mathematical Sciences, Nanyang Technological University, Singapore 637371, Singapore}
\affiliation{Complexity Institute, Nanyang Technological University, Singapore 637335, Singapore}

\author{Mile Gu}
\email{mgu@quantumcomplexity.org}
\affiliation{Nanyang Quantum Hub, School of Physical and Mathematical Sciences, Nanyang Technological University, Singapore 637371, Singapore}
\affiliation{Centre for Quantum Technologies. National University of Singapore, 3 Science Drive 2, Singapore 117543, Singapore}
\affiliation{MajuLab, CNRS-UNS-NUS-NTU International Joint Research Unit UMI 3654, Singapore}

\date{\today}


\begin{abstract}
Elementary cellular automata (ECA) present iconic examples of complex systems. Though described only by one-dimensional strings of binary cells evolving according to nearest-neighbour update rules, certain ECA rules manifest complex dynamics capable of universal computation. Yet, the classification of precisely which rules exhibit complex behaviour remains somewhat an open debate. Here we approach this question using tools from quantum stochastic modelling, where quantum statistical memory -- the memory required to model a stochastic process using a class of quantum machines -- can be used to quantify the structure of a stochastic process. By viewing ECA rules as transformations of stochastic patterns, we ask: \emph{Does an ECA generate structure as quantified by the quantum statistical memory, and can this be used to identify complex cellular automata}? We illustrate how the growth of this measure over time correctly distinguishes simple ECA from complex counterparts. Moreover, it provides a spectrum on which we can rank the complexity of ECA, by the rate at which they generate structure. 
\end{abstract}

\maketitle


\section{Introduction}

We all have some intuition of complexity. When presented with a highly ordered periodic sequence of numbers, we can often spot a simple pattern. Meanwhile, highly disordered processes, such as a particle undergoing Brownian motion, can be described using tractable mathematical models~\cite{Deutch1987, chew2002microscopic}.
Other processes -- such as stock markets, living systems, and universal computers -- lack such simple descriptions and are thus considered complex; their dynamics often lie somewhere between order and disorder~\cite{crutchfield2011between}. Yet, a quantitative criterion for identifying precisely what is complex remains challenging even in scenarios that appear deceptively simple. 

\begin{figure}
\includegraphics[width=\linewidth]{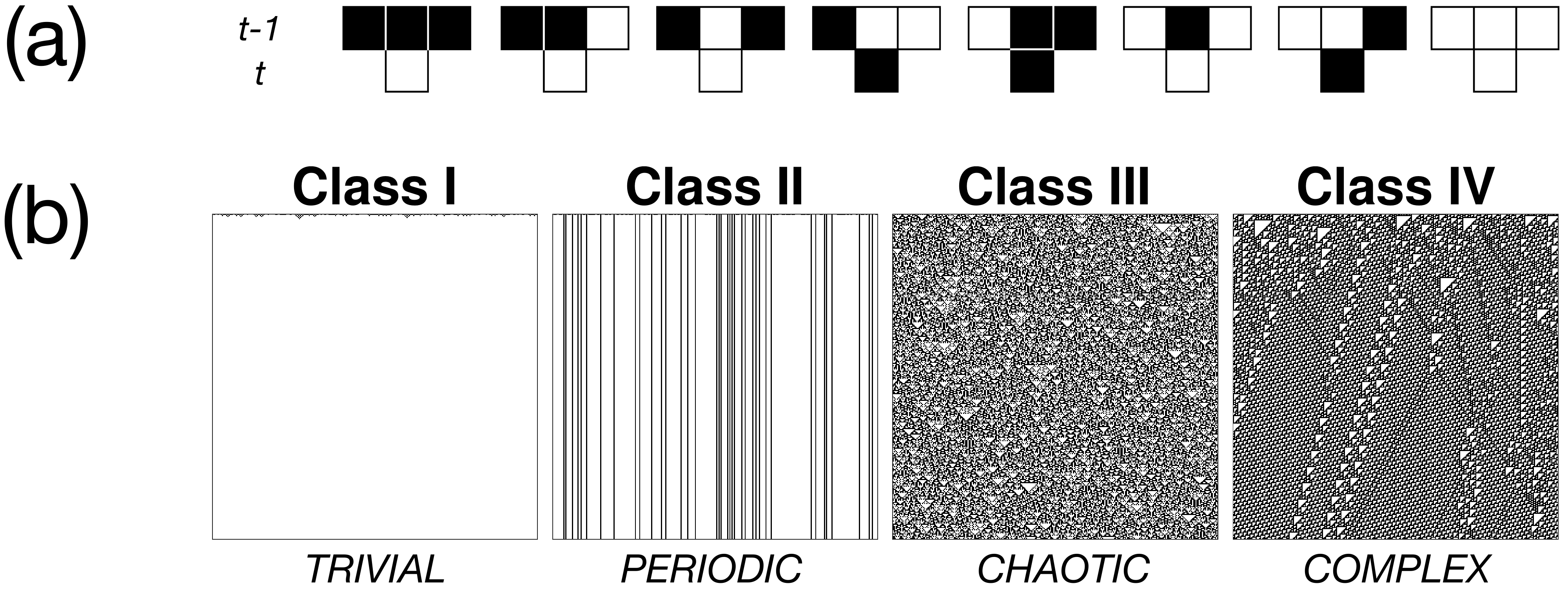}
\caption{(a) The state of each cell in an ECA takes on a binary value, and is deterministically updated at each timestep according to the current states of it and its nearest neighbours. A rule number codifies this update by converting a binary representation of the updated states to decimal. Shown here is Rule 26 $(00011010)_2$. (b) The evolution of an ECA can be visualised via a two-dimensional grid of cells, where each row corresponds to the full state of the ECA at a particular time, and columns the evolution over time. Depicted here are examples of ECA for each of Wolfram's four classes.}
\label{fig.ECA}
\end{figure}

Elementary cellular automata (ECA) provide an apt example of systems that may conceal vast complexities. Despite featuring dynamics involving only nearest-neighbour update rules on a one-dimensional chain of binary cells (see \figref{fig.ECA}), they can exhibit remarkably rich behaviour. Wolfram's initial classification of their behaviour revealed both simple ECAs that were either highly ordered (e.g., periodic dynamics) or random (e.g., chaotic dynamics), as well as others complex enough to encode universal computers~\cite{Wolfram1984a, Wolfram1984b, Cook2004}. Yet, an identification and classification of which ECA rules are complex is not as clear cut as one may expect; Wolfram's initial classification was soon met with myriad of alternative approaches~\cite{Langton1986, Li1987, Gutowitz1987, Li1990,Gutowitz1990a, Li1990a, Langton1990, Binder1993, Wuensche1999, Ninagawa2008, Ruivo2013, Martinez2013}. While they agree in extremal cases, the classification of many borderline ECA lacks a common consensus.

In parallel to these developments, there has also been much interest in quantifying and understanding the structure within stochastic processes. In this context, the statistical complexity has emerged as a popular candidate~\cite{crutchfield1989inferring, shalizi2001computational}. It asks ``\emph{How much information does a model need to store about the past of a process for statistically-faithful future prediction}?". By this measure, a completely-ordered process that generates a homogeneous sequence of $0$s has no complexity, as there is only one past and thus nothing to record. A completely random sequence also has no complexity -- all past observations lead to the same future statistics, and so a model gains nothing by tracking any past information. However, modelling general processes between these extremes -- such as those that are highly non-Markovian -- can require the tracking of immense amounts of data, and thus indicate high levels of complexity. These favourable properties, together with the clear operational interpretation, has motivated the wide use of statistical complexity as a quantifier of structure in diverse settings~\cite{Hanson1997, Goncalves1998, Park2007, Haslinger2010, huynh2015arcfractal}. 

Further developments have shown that even when modelling classical data, the most efficient models are quantum mechanical~\cite{Gu2012, mahoney2016occam, elliott2018superior, binder2018practical, elliott2019memory, liu2019optimal, loomis2019strong}. This has led to a quantum analogue of the statistical complexity -- the quantum statistical memory -- with distinct qualitative and quantitative behaviour~\cite{Suen2017, Aghamohammadi2017a, Garner2017b, Aghamohammadi2017b, elliott2018superior, elliott2019memory, thompson2018causal, elliott2020extreme, elliott2021quantum, elliott2022quantum} that demonstrates even better alignment with our intuitive notions of what is complex~\cite{Suen2017, Suen2022surveying}, and a greater degree of robustness~\cite{ho2020robust}.

Here we ask: \emph{Can these developments offer a new way to identify and classify complex ECAs}? To approach this, at each timestep $t$ of the ECA's evolution we interpret the ECA state as a stochastic string, for which we can assign a quantum statistical memory cost $C_q^{(t)}$. We can then chart the evolution of this over time, and measure how the amount of information that must be tracked to statistically replicate the ECA state grows over time. Asserting that complex ECA dynamics should yield ever more complex strings, and thus enable continual growth in structure, we propose to then interpret the growth of $C_q^{(t)}$ as a measure of an ECA's complexity. Our results indicate that this interpretation has significant merit: ECA that were unanimously considered to be complex in prior studies exhibit continual growth in $C_q^{(t)}$, while those unanimously considered simple do not. Meanwhile, its application to more ambiguous cases offers a new perspective of their relative complexities, and indicates that certain seemingly-chaotic ECA may possess some structure.


\section{Background}

\subsection{Stochastic Processes, Models, and Structure}

To formalise our approach, we first need to introduce some background on stochastic processes and measures of structure. A bi-infinite, discrete-time stochastic process is described by an infinite series of random variables $Y_i$, where index $i$ denotes the timestep. A consecutive sequence is denoted by $Y_{l:m}:=Y_lY_{l+1}\ldots Y_{m-1}$, such that if we take 0 to be the present timestep, we can delineate a past $\past{Y}:=\lim_{L\to\infty}Y_{-L:0}$ and future $\fut{Y}:=\lim_{L\to\infty}Y_{0:L}$. Associated with this is a set of corresponding variates $y_i$, drawn from a distribution $P(\past{Y},\fut{Y})$. A stationary process is one that is translationally invariant, such that $P(Y_{0:\tau}) = P(Y_{k:\tau+k})\forall\tau, k \in \mathcal{Z}$.

To quantify structure within such stochastic processes, we make use of computational mechanics~\cite{crutchfield1989inferring, shalizi2001computational} -- a branch of complexity science. Consider a model of a stochastic process that uses information from the past to produce statistically-faithful future outputs. That is, for any given past $\past{y}$, the model must produce a future $\fut{y}$, one step at a time, according to the statistics of $P(\fut{Y}|\past{y})$. Since storing the entire past is untenable, operationally, this requires a systematic means of encoding each $\past{y}$ into a corresponding memory state $S_{\past{y}}$, such that the model can use its memory to produce outputs according to $P(\fut{Y}|S_{\past{y}})=P(\fut{Y}|\past{y})$. Computational mechanics then ascribes the complexity of the process to be the memory cost of the simplest model, i.e., the smallest amount of information a model must store about the past to produce statistically-faithful future outputs. This memory cost is named the \emph{statistical complexity} $C_\mu$.

In the space of classical models, this minimal memory cost is achieved by the $\varepsilon$-machine of the process. They are determined by means of an equivalence relation $\past{y}\sim\past{y}'\iff P(\fut{Y}|\past{y}) = P(\fut{Y}|\past{y}')$, equating different pasts iff they have coinciding future statistics. This partitions the set of all pasts into a collection of equivalence classes $\mathcal{S}$, called causal states. An $\varepsilon$-machine then operates with memory states in one-to-one correspondence with these equivalence classes, with an encoding function $\varepsilon$ that maps each past to a corresponding causal state $S_j=\varepsilon(\past{y})$. The resulting memory cost is given by the Shannon entropy over the stationary distribution of causal states:
\begin{equation}
C_\mu = H[P(S_j)] = -\sum_j P(S_j) \log_2 P(S_j),
\end{equation}
where $P(S_j)=\sum_{\past{y}|\varepsilon(\past{y})=S_j}P(\past{y})$.

\begin{figure}
\includegraphics[width=\linewidth]{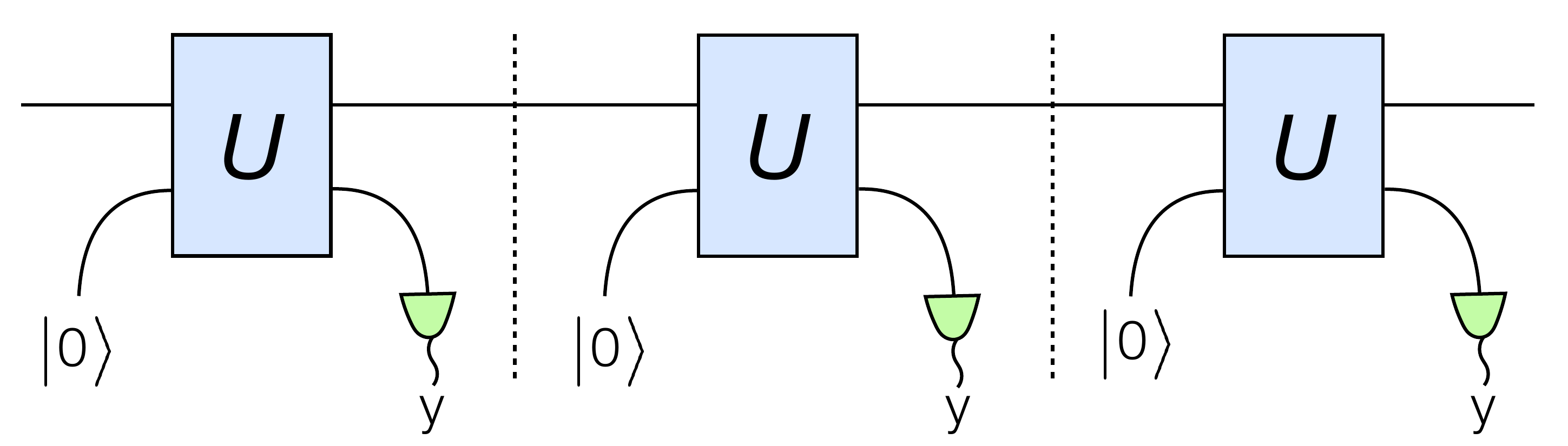}
\caption{A quantum model consists of a unitary operator $U$ acting on a memory state $\ket{\sigma_j}$ and blank ancilla $\ket{0}$. Measurement of the ancilla produces the output symbol, with the statistics of the modelled process realised through the measurement statistics.}
\label{fig.quantummodels}
\end{figure}

When the model is quantum mechanical, further reduction of the memory cost is possible, by mapping causal states to non-orthogonal quantum memory states $S_j\to\ket{\sigma_j}$. These quantum memory states are defined implicitly through an evolution operator, such that the action of the model at each timestep is to produce output $y$ with probability $P(y|S_j)$, and update the memory state~\cite{binder2018practical, liu2019optimal}. Specifically, this evolution is given by 
\begin{equation}
\label{eq.quantumcausalstates}
U\ket{\sigma_j}\ket{0} = \sum_{y} \sqrt{P(y|S_j)} \ket{\sigma_{\lambda(y,j)}} \ket{y}
\end{equation}
where $U$ is a unitary operator also implicitly defined by this expression. Here, $\lambda(y,j)$ is a deterministic update function that updates the memory state to that corresponding to the causal state of the updated past. Sequential application of $U$ then replicates the desired statistics (see Fig. \ref{fig.quantummodels}).

The memory cost of such quantum models is referred to as the \emph{quantum statistical memory}~\footnote{The term `quantum statistical memory' is used in place of `quantum statistical complexity' as such quantum machines may not necessarily be memory-minimal among all quantum models~\cite{Suen2017, liu2019optimal}.}. Paralleling its classical counterpart, this is given by the von Neumann entropy of the steady-state of the quantum model's memory $\rho = \sum_j P(S_j) \ket{\sigma_j}\bra{\sigma_j}$:
\begin{equation}
C_q = -\text{Tr}(\rho \log_2 \rho).
\end{equation}
The non-orthogonality of the quantum memory states ensures that in general $C_q < C_\mu$~\cite{Gu2012}, signifying that the minimal past information needed for generating future statistics -- if all methods of information processing are allowed -- is generally lower than $C_\mu$. This motivates $C_q$ as an alternative means of quantifying structure~\cite{Gu2012, tan2014towards, Suen2017, Aghamohammadi2017a}, where it has shown stronger agreement with intuitions of complexity~\cite{Suen2017,Suen2022surveying}. In addition to this conceptual relevance, the continuity of the von Neumann entropy makes $C_q$ much more well-behaved compared to $C_\mu$, such that a small perturbation in the underlying stochastic process leads to only a small perturbation in $C_q$, which becomes particularly relevant when inferring complexity from a finite sample of a process~\cite{ho2020robust}. 

Since a quantum model can be systematically constructed from the $\varepsilon$-machine of a process, a quantum model and associated $C_q$ can be inferred from data by first inferring the $\varepsilon$-machine. However, the quantum model will then inherit errors associated with the classical inference method, such as erroneous pairing or separation of pasts into causal states. For this reason, a quantum-specific inference protocol was recently developed~\cite{ho2020robust} that bypasses the need to first construct an $\varepsilon$-machine, thus circumventing some of these errors. It functions by scanning through the stochastic process in moving windows of size $L+1$, in order to estimate the probabilities $P(Y_{0:L+1})$, from which the marginal and conditional distributions $P(Y_{0:L})$ and $P(Y_0|Y_{-L:0})$ can be determined. From these, we construct a set of inferred quantum memory states $\{\ket{\varsigma_{y_{-L:0}}}\}$, satisfying
\begin{equation}
\label{eq.quantuminferenceprotocol}
U \ket{\varsigma_{y_{-L:0}}} \ket{0} = \sum_{y_0 } \sqrt{P(y_0 | y_{-L:0})}  \ket{\varsigma_{y_{-L+1:1}}} \ket{y_0}.
\end{equation}
for some suitable unitary operator $U$. When $L$ is greater than or equal to the Markov order of the process, and the probabilities used are exact, this recovers the same quantum memory states as the exact quantum model Eq.~\eqref{eq.quantumcausalstates}, where the quantum memory states associated to two different pasts are identical iff the pasts belong to the same causal state. Otherwise, if $L$ is sufficiently long to provide a `good enough' proxy for the Markov order, and the data stream is long enough for accurate estimation of the $L+1$-length sequence probabilities, then the quantum model will still be a strong approximation with a similar memory cost. From the steady-state of these inferred quantum memory states, the quantum statistical memory $C_q$ can be inferred~\cite{ho2020robust}.

However, the explicit quantum model need not be constructed as part of the inference of the quantum statistical memory. The spectrum of the quantum model steady-state is identical to that of its Gram matrix~\cite{Horn2012}. For the inferred quantum model, this Gram matrix is given by
\begin{equation}
\label{eq.grammatrix}
    G_{y_{-L:0} y'_{-L:0}} = \sqrt{ P(y_{-L:0}) P(y'_{-L:0}) } \sum_{y_{0:L}} \sqrt{P(y_{0:L}|y_{-L:0}) P(y_{0:L}|y'_{-L:0})}.
\end{equation}
The associated conditional probabilities $P(Y_{0:L}|Y_{-L:0})$ can either be estimated from compiling the $P(Y_0|Y_{-L:0})$ using $L$ as a proxy for the Markov order, or directly by frequency counting of strings of length of $2L$ in the data stream. Then, the quantum inference protocol yields an estimated quantum statistical memory $\tilde{C}_q$:
\begin{equation}
    \tilde{C}_q = -\text{Tr} (G \log_2 G).
\end{equation}


\subsection{Classifying Cellular Automata}

The state of an ECA can be represented as an infinite one-dimensional chain of binary cells that evolve dynamically in time. At timestep $t$, the states of the cells are given by $x_i^t \in \mathcal{A} =  \{0,1\}$, where $i$ is the spatial index. Between each timestep, the states of each cell evolve synchronously according to a local update rule $x_{i}^{t} = \mathcal{F}(x^{t-1}_{i-1}, x^{t-1}_{i}, x^{t-1}_{i+1})$. There are $2^{2^3} = 256$ different possible such rules, each defining a different ECA~\cite{Wolfram1984a}. A rule number is specified by converting the string of possible outcomes for each past configuration from binary into decimal as illustrated in \figref{fig.ECA}(a). After accounting for symmetries (relabelling $0$ and $1$, and mirror images), $88$ independent rules remain. Each of these rules can yield very different behaviour, motivating their grouping into classes. One such popular classification is that of Wolfram~\cite{Wolfram1984a, Wolfram1984b}, describing four distinct classes (see \figref{fig.ECA}(b)):
\begin{description}
\setlength\itemsep{-0.3em}
    \item[Class I]\emph{Trivial.} The ECA evolves to a uniform state.
    \item[Class II]\emph{Periodic.} The ECA evolves into a stable or periodic state.
    \item[Class III]\emph{Chaotic.} The ECA evolves into a seemingly random state with no discernable structure.
    \item[Class IV]\emph{Complex.} The ECA forms structures in its state that can interact with each other.
\end{description}
Classes I, II and III are all considered simple, in the sense their future behaviour is statistically easy to describe. However, Class IV ECAs are deemed complex, as they enable highly non-trivial forms of information processing, including some capable of universal computation~\cite{Cook2004}.

Consider for example, two extremal cases that typify simplicity and complexity respectively:
\begin{description}
    \item[\textbf{Rule 30}] Used for generating pseudo-random numbers~\cite{Wolfram1986a, Rule30RNG}, it is an iconic example of a chaotic Class III ECA.
    \item[\textbf{Rule 110}] Proven capable of universal computation by encoding information into `gliders'~\cite{Cook2004}, it is the iconic example of a complex Class IV ECA.
\end{description}

Yet the boundaries between classes, especially that of Classes III and IV, still lacks universal consensus. This has motivated several diverse methods to better classify ECA, including the analysis of power spectra~\cite{Li1987, Ninagawa2008, Ruivo2013}, Langton parameters~\cite{Langton1986, Li1990, Li1990a, Langton1990, Binder1993}, filtering of space-time dynamics~\cite{Wuensche1999}, hierarchical classification~\cite{Gutowitz1990a}, computing the topology of the Cantor space of ECA~\cite{schule2012topologicalECA, kurka1997languages, kurka2001topological, gilman1987classes}, and mean field theory~\cite{Gutowitz1987}. These schemes feature discrete classes but do not yield identical conclusions, highlighting the inherent difficulties in discerning complexity and randomness.

Illustrative examples of ECA with ambiguous classification include:
\begin{description}
    \item[\textbf{Rule 54}] This rule exhibits interacting glider systems like Rule 110~\cite{Hanson1997,Martin2000,Martinez2014}, but its capacity for universal computation is unknown.
    \item[\textbf{Rule 18}] Assigned by Wolfram to Class III, though subsequent studies indicate it contains anomalies known as kinks that can propagate and annihilate other kinks~\cite{Grassberger1984, Eloranta1992}. This indicates some level of structure.
\end{description}

Ambiguous rules such as these complicate attempts that seek to determine a border between simple and complex ECA. Rather, we may instead envision trying to place such ECA on a spectrum, with one end having ECA that are clearly in Class III, and the other ECA clearly in Class IV, with the more ambigious cases on a gradated scale in between. To our knowledge, ours is the first approach that refines ECA classification into such a hybrid mix of discrete classes with a continuous spectrum.


\section{A Stochastic Perspective}

\subsection{Overview and Analytical Statements}

We will use the tools of computational mechanics to classify ECA, placing them on such a spectrum as described above. To do so, we first need to describe ECA in the language of stochastic processes. Observe that if we initialise an ECA at random, then the state of the ECA at $t = 0$ is described by a spatially-stationary stochastic process  $\pastfut{Y}^{(0)}$  -- specifically, a completely random process which has $C_q^{(0)} = 0$. The evolution of the ECA transforms this into a new stochastic process; let $\pastfut{Y}^{(t)}$ describe the stochastic process associated with the ECA state at timestep $t$, where the spatial index of the state takes the place of the temporal index in the process. As the update rules for ECA are translationally invariant, the $\pastfut{Y}^{(t)}$ are also spatially-stationary, and thus possess a well-defined complexity measure $C_q^{(t)}$. We emphasise that the evolution of the ECA remains deterministic; the stochasticity in the processes extracted from the ECA state emerges from a scanning of successive cells in the ECA state at a given point in time without knowledge of any prior state of the ECA (including its initial, random, state).

Charting the evolution of $C_q(t)$ with successive applications of the update rule can then help us quantify the amount of structure created or destroyed during the evolution of an ECA. We propose the following criterion:\\

\noindent \emph{An ECA is complex if $C_q^{(t)}$ grows with $t$ without bound.}\\

The rationale is that complex dynamics should be capable of generating new structure. Simple ECA are thus those that generate little structure, and correspondingly have $C_q^{(t)}$ stagnate. At the other extreme, ECA capable of universal computation can build up complex correlations between cells arbitrarily far apart as $t$ grows, requiring us to track ever-more information; thus, $C_q^{(t)}$ should grow.

We can immediately make the following statements regarding all ECA in Classes I and II, and some in Class III -- highlighting that our criterion also identifies simple ECA.
 
\begin{itemize}
    \item[(1)] $\lim_{t\to\infty}C^{(t)}_q = 0$ for all Class I ECA.
    \item[(2)] $C^{(t)}_q$ must be asymptotically bounded for all Class II ECA. That is, there exists a sufficiently large $T$ and constant $K$ such that $C^{(t)}_q \leq K$ for all $t > T$.
    \item[(3)] $C^{(t)}_q\approx 0$ for ECA suitable for use as near-perfect pseudo-random number generators.
\end{itemize}

Statement (1) follows from the definition that Class I ECA are those that evolve to homogeneous states; at sufficiently large $t$, $\pastfut{y}^{(t)}$ is a uniform sequence of either all $0$s or all $1$s, for which $C^{(t)}_q = 0$ (since there is only a single causal state).
With Class II ECA defined as those that are eventually periodic with some periodicity $\tau$ (modulo spatial translation, which does not alter $C_q$), this implies that their $C_q^{(t)}$ similarly oscillates with at most the same periodicity $\tau$, with some maximal value bounded by some constant $K$, thus statement (2) follows. Finally to see statement (3), we note that a perfect random number generator should have no correlations or structure in their outputs, and hence have $C^{(t)}_q = 0$; the continuity of $C_q$ then ensures that near-perfect generators should have $C^{(t)}_q\approx0$.


\subsection{Numerical Methodology}

\begin{figure}
\includegraphics[width=\linewidth]{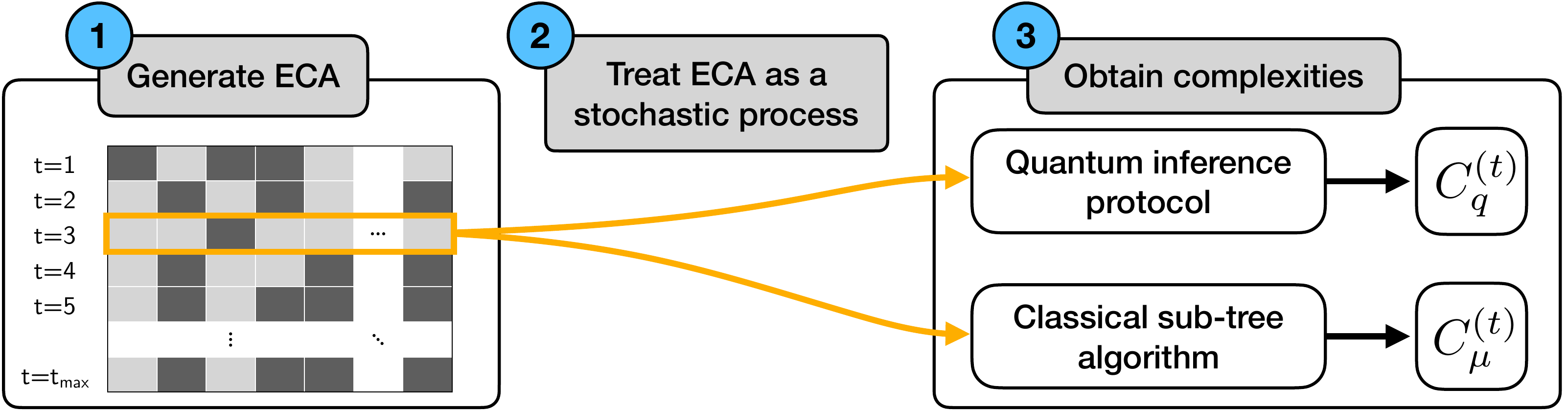}
\caption{An ECA is evolved from random initial conditions. Treating the ECA state at each timestep as a stochastic process, we then infer the quantum statistical memory $C^{(t)}_q$ and classical statistical complexity $C^{(t)}_\mu$. By observing how these measures change over time, we are able to deduce the complexity of the ECA rule.}
\label{fig.workflow}
\end{figure}

For the ECA that lie between Class III and IV of interest to us here, we must deal with their potential computational irreducibility. There are thus generally no shortcuts to directly estimate the long-time behaviour of $C^{(t)}_q$ in such ECA. We therefore employ numerical simulation. We approximate each ECA of interest with a finite counterpart, where each time-step consists of $W=64,000$ cells, such that the state of the ECA at time $t$ is described by a finite data sequence $y_{0:W}^{(t)}$, up to a maximum time $t_\text{max} = 10^3$~\footnote{Note that this is many, many orders of magnitude smaller than the time for which a typical finite-width ECA is guaranteed to cycle through already-visited states ($\mathcal{O}(2^W)$)~\cite{Grassberger1986a}.}. We can then infer the quantum statistical memory $C^{(t)}_q$ of this sequence, using the quantum inference protocol discussed above~\cite{ho2020robust}, with $L=6$ for the history length. The workflow is illustrated in \figref{fig.workflow}.

\begin{figure}[!h]
\includegraphics[width=\linewidth]{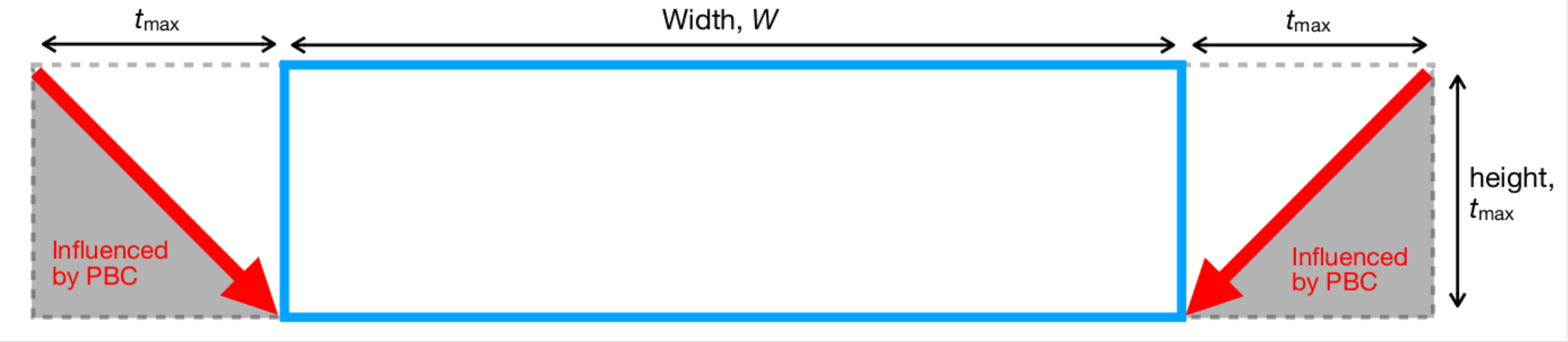}
\caption{Generation of finite-width ECA evolution with open boundaries via extended ECA with periodic boundaries.}
	\label{fig.boundary}
\end{figure}

For each ECA rule we generate an initial state for $t=1$ where each cell is randomly assigned $0$ or $1$ with equal probability, and then evolve for $t_\text{max}$ steps. We then apply the inference methods to the states at $t=1,2,3,...,9,10,20,...,90,100,200,...,t_\text{max}$; evaluating at every timestep shows little qualitative difference beyond highlighting the short periodicity of some Class II rules. We repeat five times for each rule, and determine the mean and standard deviation of $C_q^{(t)}$. To avoid boundary effects from the edges of the ECA, to obtain an ECA state of width $W$ for up to $t_{\text{max}}$ timesteps we generate an extended ECA of width $W'=W+2t_{\text{max}}$ with periodic boundary conditions and keep only the centremost $W$ cells; this is equivalent to generating a width $W$ ECA with open boundaries (see \figref{fig.boundary}). Note however that the choice of boundary condition showed little quantitative effect upon our results.

\begin{figure}[!h]
    \centering
    \includegraphics[width=\linewidth]{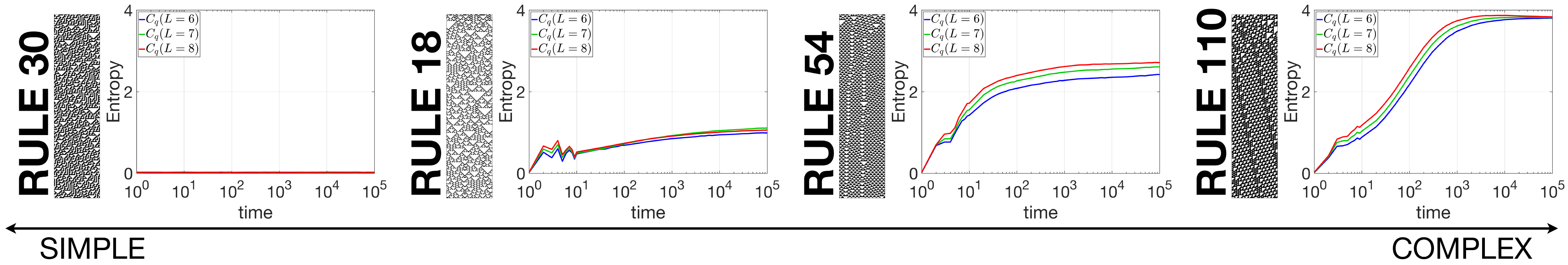}
    \caption{$C_q^{(t)}$ plots for a selection of rules with longer $L$ and larger $t_{\text{max}}$. Plots shown for $(W=64,000, L=6)$, \mbox{$(W=128,000, L=7)$}, and $(W=256,000, L=8)$.}
    \label{fig.longer}
\end{figure}

In \figref{fig.longer}) we present plots showing little difference in the qualitative features of interest with extensions up to $L=8$ and $t_{\text{max}}=10^5$, supporting our choice for these values. The exception to this is Rule 110, which appears to plateau at longer times. We believe this to be attributable to the finite width of the ECA studied -- as there are a finite number of gliders generated by the initial configuration, over time as the gliders annihilate there will be fewer of them to interact and propagate further correlations.

For completeness, we also perform analogous inference for the classical statistical complexity $C^{(t)}_\mu$ using the sub-tree reconstruction method. See Appendix A for details.

\begin{figure}[!h]
    \centering
    \includegraphics[width=\linewidth]{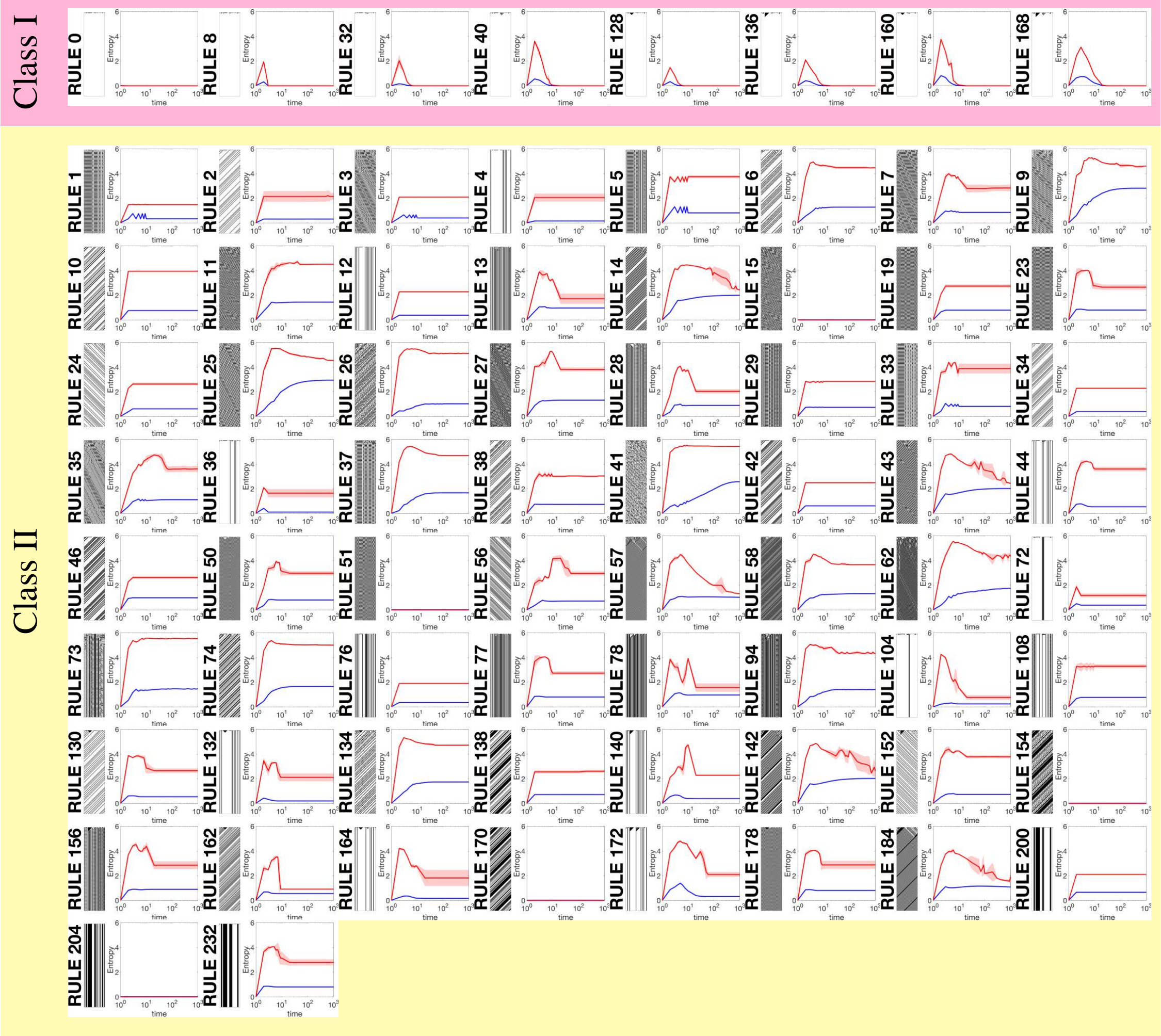}
    \caption{Evolution of $C_q^{(t)}$ (blue) and $C_\mu^{(t)}$ (red) for all Wolfram Class I and II rules. Lines indicate mean values over five different intial random states, and the translucent surrounding the standard deviation.}
    \label{fig.Class1_Class2}
\end{figure}
 
 \begin{figure}[!h]
    \centering
    \includegraphics[width=\linewidth]{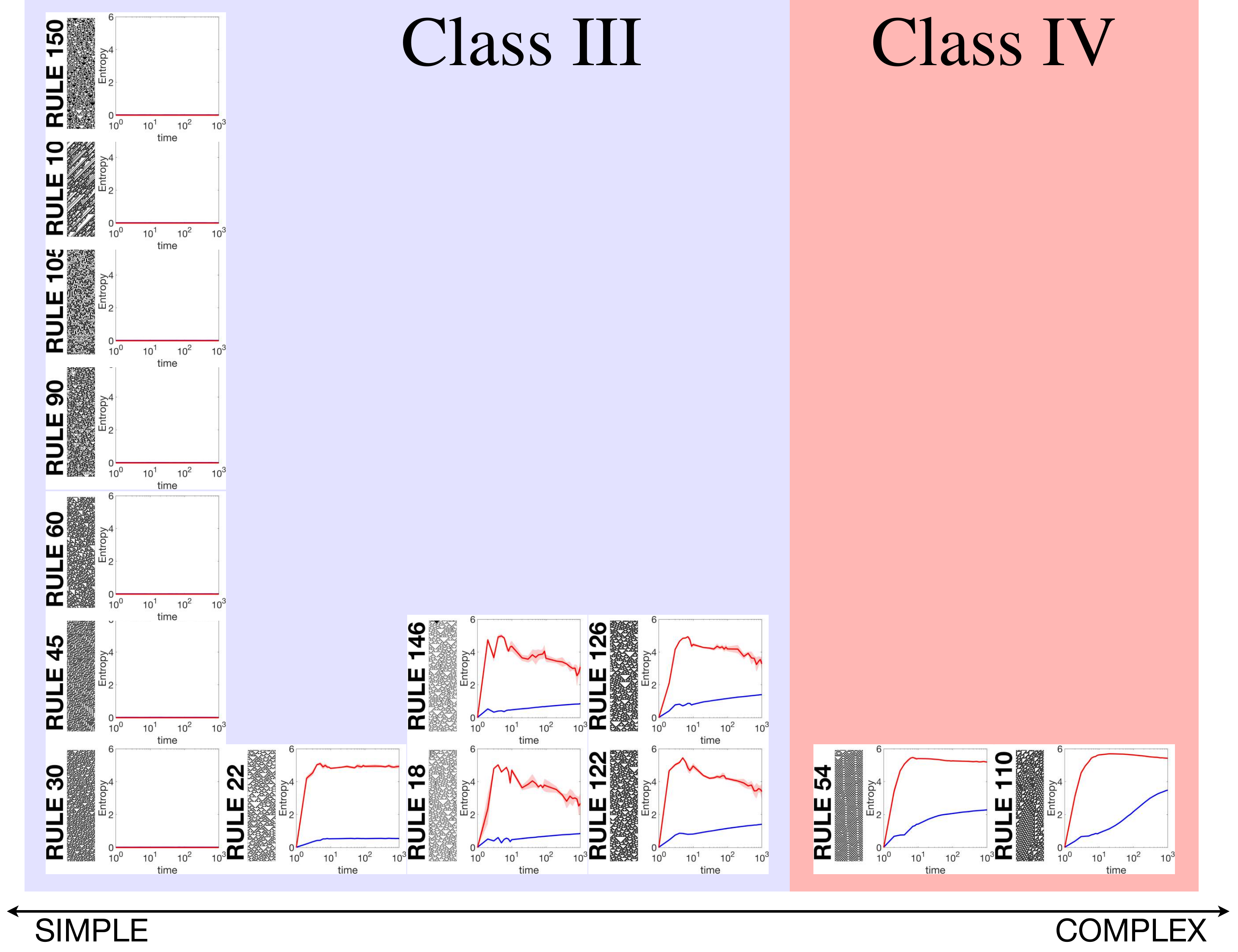}
    \caption{Evolution of $C_q^{(t)}$ (blue) and $C_\mu^{(t)}$ (red) for all Wolfram Class III and IV rules. Rules are placed on a simplicity-complexity spectrum according to the growth of $C_q^{(t)}$ with Rule 30 (Class III, simple) and Rule 110 (Class IV, complex) at the extremes. Lines indicate mean values over five different intial random states, and the translucent surrounding the standard deviation.}
    \label{fig.Class3_Class4}
\end{figure}
 
\subsection{Numerical Results}

Our results are plotted in \figref{fig.Class1_Class2} and \figref{fig.Class3_Class4}, charting the evolution of $C_q^{(t)}$ and $C_\mu^{(t)}$ for all unique ECA rules. \figref{fig.Class1_Class2} shows all ECA rules that belong to Wolfram Classes I and II; we see that $C_q^{(t)}$ indeed displays the features discussed above, i.e., that $\lim_{t\to\infty}C^{(t)}_q = 0$ for Class I ECA, and tends to a bounded value for Class II. Wolfram Class III and IV rules are displayed in \figref{fig.Class3_Class4}, where they are ranked on a simplicity-complexity spectrum according to the rate of growth of $C^{(t)}_q$.

We first observe that our extremal cases indeed sit at the extremes of this spectrum:
\begin{description}
    \item[\textbf{Rule 30}] Consistent with its role as a pseudo-random number generator, Rule 30 generates no discernable structure, yielding negligible $C^{(t)}_q \approx 0$.
    \item[\textbf{Rule 110}] Clearly exhibits the fastest growth in $C^{(t)}_q$. This aligns with its capability for universal computation; Rule 110 is able to propagate correlations over arbitrarily large distances.
\end{description}
More interesting are the rules with a more ambiguous classification. We make the following observations of illustrative examples, listed in order of increasing growth rates on $C^{(t)}_q$.
\begin{description}
    \item[\textbf{Rule 22}] We see that $C^{(t)}_q$ stagnates after a short period of initial growth, and thus behaves as a simple CA according to our complexity criterion. Indeed, prior studies of this rule suggests that it behaves as a mix of a random and a periodic process, generating a finite amount of local structure that does not propagate with time, and is thus very unlikely to be capable of universal computation~\cite{Grassberger1986a}. 
    \item[\textbf{Rule 18}] $C^{(t)}_q$ grows slowly (at around a quarter the rate of Rule 110), suggesting the presence of some complex dynamics within despite a Class III classification. Indeed, the presence of kinks in Rule 18's dynamics supports this observation.
    \item[\textbf{Rule 122}] $C^{(t)}_q$ shows similar growth to Rule 18, though slightly faster.
    \item[\textbf{Rule 54}] Aside from Rule 110, Rule 54 is the only other Wolfram Class IV rule. Consistent with this, we see that it exhibits a fast growth in $C^{(t)}_q$, likely due to its glider system propagating correlations.
\end{description}
Thus among these examples, all except Rule 22 still feature some degree of growth in $C^{(t)}_q$. We conclude that these ECA, despite their Class III classification, do feature some underlying complex dynamics. Moreover, we see that our spectrum of the relative growth rates of $C^{(t)}_q$ appears to offer a suitable means of ranking the relative complexity of each ECA. Particularly, our spectrum is consistent with the division into Class III and IV, but provides a further nuance than the traditional discrete classes. 

We remark that on the other hand, the classical $C_\mu^{(t)}$ does not appear to provide such a clear indicator of complexity in the ECA, with some Class II-IV rules even showing a reduction in complexity over time. Moreover, the instability of the measure is evident. Together, this highlights the advantages of using the quantum measure $C_q^{(t)}$ to quantify complexity, over its classical counterpart.

We further remark that while $C_q^{(t)}$ has performed admirably for this purpose, it may not be unique in this regard. There exist a multitude of measures of complexity, of varying utility and areas of application. It is of course not possible for us to explore all such possibilities, but we will briefly here comment on some of the more commonly-used measures. Perhaps the most famous measure of `complexity' is the algorithmic information~\cite{kolmogorov1963tables, solomonoff1964formal, chaitin1969simplicity}, that quantifies the shortest programme required to output a particular string. However, this is now recognised to be better suited as a measure of randomness rather than complexity, and is moreover, generically uncomputable. Similarly, the thermodynamic depth~\cite{lloyd1988complexity} has been linked with randomness rather than complexity, and suffers from ambiguities in its proper calculation~\cite{crutchfield1999thermodynamic}. However, one potentially suitable alternative is the excess entropy $E$, that quantifies the mutual information between the past and future outputs of a stochastic process~\cite{shalizi2001computational}, and is a lower bound on $C_q$ and $C_\mu$. We see from \figref{fig.CmuCqEE_moreL} that $E$ may also work in place of $C_q$ here as a suitable complexity measure to identify complex ECA. Indeed, qualitative similarities between $C_q$ and $E$ have previously been noted~\cite{Suen2022surveying}. We leave a more complete analysis of the qualitative similarities and difference of $C_q$ and $E$ -- in this context and beyond -- as a question for future work.


\section{Discussion}

Here, we introduced new methods for probing the complexity of ECA. By viewing the dynamics of a one-dimensional cellular automata at each time-step as a map from one stochastic process to another, we are able to quantify the structure of the ECA state at each timestep. Then, by initialising an ECA according to a random process with no structure, we can observe if the ECA's update rules are able to transduce to stochastic processes with increasing structure. In this picture an ECA is considered simple if the structure saturates to a bounded value over time, and is complex if it exhibits continued growth. To formalise this approach, we drew upon computational mechanics, which provides a formal means of quantifying structure within stochastic process as the memory cost needed to simulate them. We found that the memory cost associated with quantum models -- the quantum statistical memory $C_q$ -- performed admirably at identifying complex ECA. It provides agreement with prior literature when there is concensus on the ECA's complexity, and is able to place the remaining ECA of ambiguous complexity on a simplicity-complexity spectrum.

One curiosity of this proposal is the unlikely juxtaposition of using a measure of complexity involving quantum information, when the systems involved (ECA) are purely classical objects. One rationale would be mathematical practicality -- the continuity properties of $C_q$ make it more stable, while its slower scaling makes it less susceptible to saturating numerical limitations. On the other hand, the use of quantum measures for classical objects is not entirely alien; quantum computers are expected to solve certain classical computational problems more efficiently than classical computers, and one may thence argue that the true complexity of a process should account for all physical means of its simulation.

Our results open some interesting queries. Several of the ECA our methodology identified as complex lie within Wolfram Class III, suggesting that some ECA many considered to feature only randomness may actually be capable of more complex information processing tasks. However, for only one of these (Rule 18) was a means to do this previous identified~\cite{Grassberger1984, Eloranta1992}. Could the other ECA with equal or higher $C^{(t)}_q$ growth, such as Rule 122, also feature such dynamics when properly filtered? More generally, it appears intuitive that universal computation and continual $C_q^{(t)}$ growth should be related, but can such a relation be formalised? Finally, while our studies here have focused on ECA, the methdology naturally generalises to any one-dimensional cellular automata -- such as those with more cell states, longer-range update rules, asynchronous tuning~\cite{Uragami2018}, and memory~\cite{Martinez2010}; it would certainly be interesting to see if our framework can provide new insight into what is complex in these less-extensively studied situations.

\acknowledgements 
This work was funded by the National Research Foundation, Singapore, and Agency for Science, Technology and Research (A*STAR) under its QEP2.0 programme (NRF2021-606 QEP2-02-P06), the Singapore Ministry of Education Tier grants RG146/20 and RG77/22, grant FQXi-RFP from the Foundational Questions Institute and Fetzer Franklin Fund (a donor advised fund of the Silicon Valley Community Foundation), and the University of Manchester Dame Kathleen Ollerenshaw Fellowship.

\section*{DATA AVAILABILITY}
The datasets generated during and/or analysed during the current study are available from the corresponding author on reasonable request.


\appendix

\section{Sub-tree reconstruction algorithm}

Here, inference of the classical statistical complexity $C_\mu$ is achieved through the sub-tree reconstruction algorithm~\cite{crutchfield1989inferring}. It works by explicitly building an $\varepsilon$-machine of a stochastic process, from which $C_\mu$ may readily be deduced. The steps are detailed below.

\begin{figure*}[!h]
\includegraphics[width=0.9\linewidth]{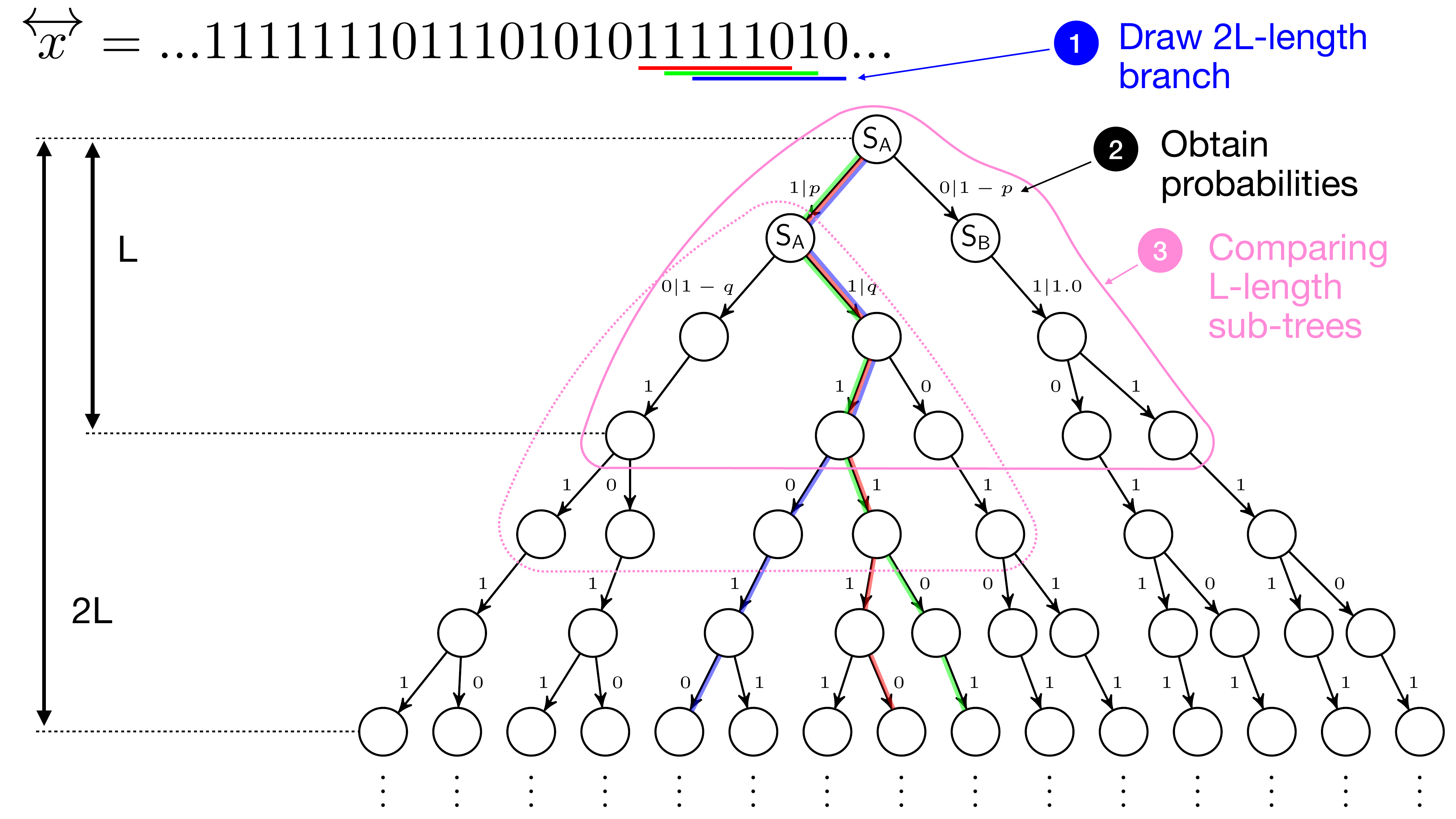}
\caption{The sub-tree reconstruction algorithm, here illustrated for $L=3$.}
\label{fig.subtree}
\end{figure*}

\textbf{1. Constructing a tree structure.}
The sub-tree construction begins by drawing a blank node to signify the start of the process with outputs $y \in \mathcal{A}$. A moving window of size $2L$ is chosen to parse through the process.
Starting from the blank node, $2L$ successive nodes are created with a directed link for every $y$ in each moving window $\{y_{0:2L}\}$. For any sequence starting from $y_0$ within $\{y_{0:2L}\}$ whose path can be traced with existing directed links and nodes, no new links and nodes are added. New nodes with directed links are added only when the $\{y_{0:2L}\}$ does not have an existing path. This is illustrated in \figref{fig.subtree}

For example, suppose $y_{0:6} = 000000$ , giving rise to six nodes that branch outwards in serial from the initial blank node. If $y_{1:7} = 000001$, the first five nodes gain no new branches, while the sixth node gains a new branch connecting to a new node with a directed link. Each different element of $|\mathcal{A}|^{2L}$ has its individual set of directed links and nodes, allowing a maximum of $|\mathcal{A}|^{2L}$ branches that originate from the blank node.

\textbf{2. Assigning probabilities.}
The probability for each branch from the first node to occur can be determined by the ratio of the number of occurences the associated strings to the total number of strings. Correspondingly, this allows each link to be denoted with an output $y$ with its respective transition probability $p$.

\vspace{0.4cm}

\textbf{3. Sub-tree comparison.}
Next, starting from the initial node, the tree structure of $L$ outputs is compared against all other nodes. Working through all reachable $L$ nodes from the initial node, any nodes with identical $y|p$ and branch structure of $L$ size are given the same label. Because of finite data and finite $L$, a $\chi^2$ test is used to account for statistical artefacts. The $\chi^2$ test will merge nodes that have similar-enough tree structures. This step essentially enforces the causal equivalence relation on the nodes.

\vspace{0.4cm}

\textbf{4. Constructing the $\varepsilon$-machine.}
It is now possible to analyse each individually-labelled node with their single output and transition probability to the next node. An edge-emitting hidden Markov model of the process can then be drawn up.
This edge-emitting hidden Markov model represents the (inferred) $\varepsilon$-machine of the process.

\vspace{0.4cm}

\textbf{5. Computing the statistical complexity.}
The hidden Markov model associated with the $\varepsilon$-machine has a transition matrix $T_{kj}^y$ giving the probability of the next output being $y$ given we are in causal state $S_j$, and $S_k$ being the causal state of the updated past. The steady-state of this (i.e., the eigenvector $\pi$ satisfying $\sum_yT^y\pi=\pi$) gives the steady-state probabilities of the causal states. Taking $P(S_j)=\pi_j$, we then have the Shannon entropy of this distribution gives the statistical complexity:
\begin{equation}
	C_\mu := H[ P(S_j) ] = -\sum_j P(S_j) \log_2 P(S_j).
\end{equation}

For parity with the quantum inference protocol, we used $L=6$ when inferring the $C_\mu^{(t)}$ of the ECA states, and set the tolerance of the $\chi^2$ test to 0.05.

\section{Convergence with $L$}

\begin{figure}[h]
    \centering
    \includegraphics[width=1.0\linewidth]{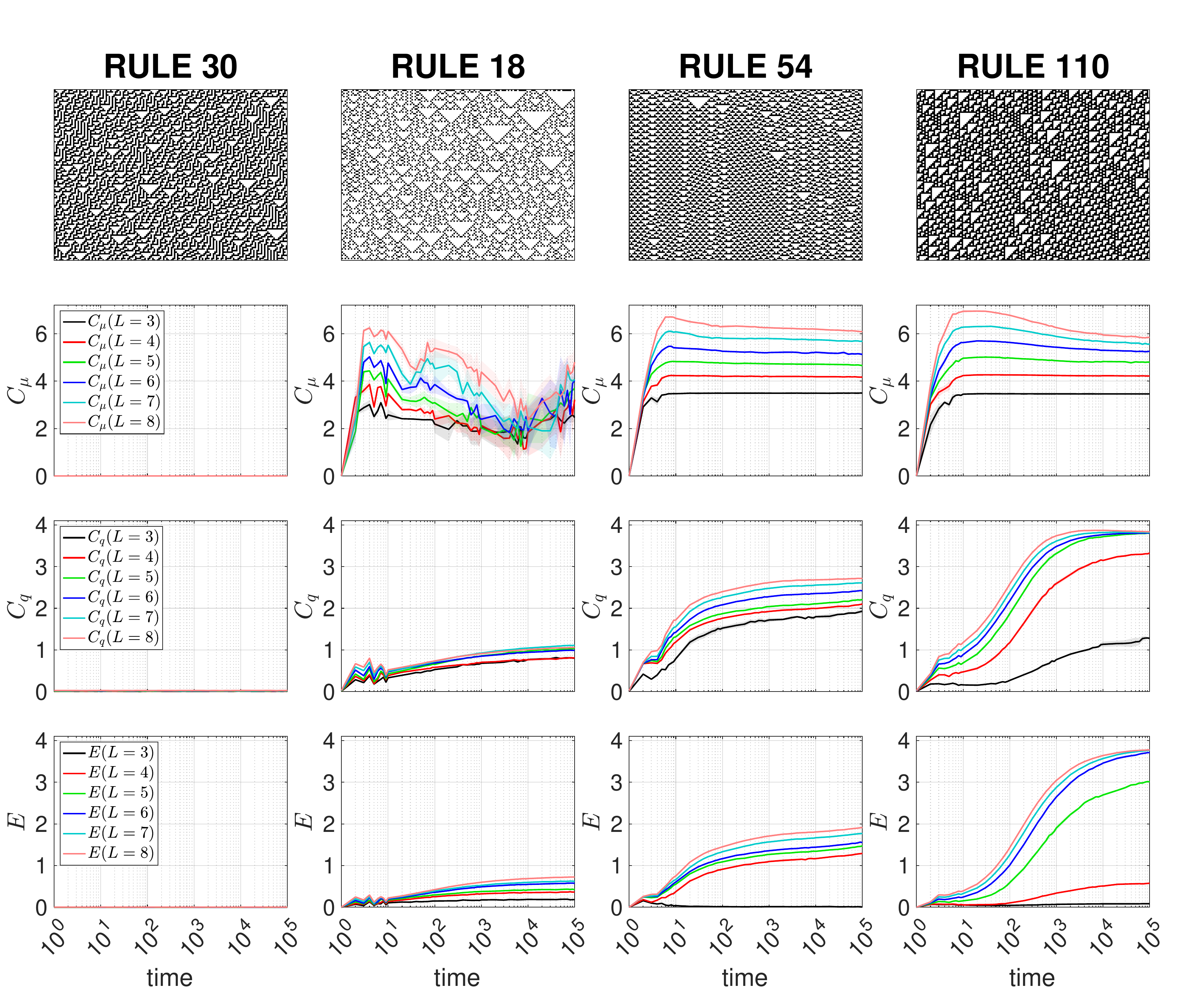}
    \caption{Growth of $L$-length ($L=3:8)$ estimates of $C_\mu$, $C_q$, and $E$ with time, up to a maximum of $t_\text{max} = 10^5$ time steps. Different $L$ correspond to different widths $W=10^3*2^L$.}
    \label{fig.CmuCqEE_moreL}
\end{figure}

We here provide an expanded form of \figref{fig.longer} in \figref{fig.CmuCqEE_moreL} regarding the convergence of our measures of complexity with increasing probability window lengths $L$, now including $L=3:8$, and also showing analogous plots for $C_\mu$ and $E$ alongside $C_q$. We note that the excess entropy $E$ is estimated for finite length windows using $E(L)=LH(X_{0:L-1})-(L-1)H(X_{0:L})$. 

\clearpage

\bibliography{ref}

\end{document}